\documentclass[twoside,a4paper,11pt]{sea22}
\usepackage{graphicx}
\usepackage{hyperref}
\usepackage{media9}
\begin{document}
\pagenumbering{arabic}
\pagestyle{myheadings}
\thispagestyle{empty}
{\flushleft\includegraphics[width=\textwidth,bb=58 650 590 680]{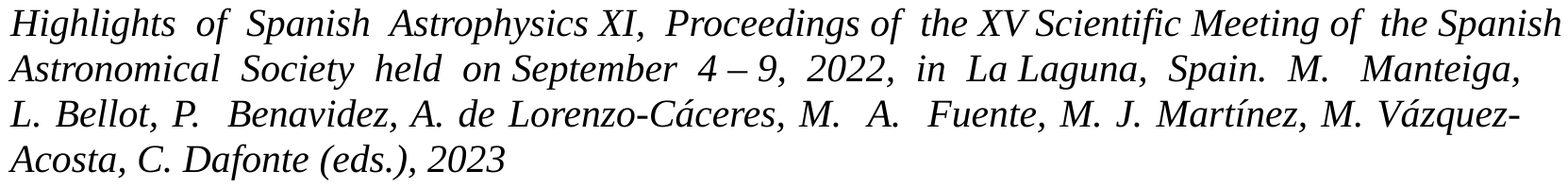}}
\vspace*{0.2cm}
\begin{flushleft}
{\bf {\LARGE
%
The upcoming spectroscopic powerhouses at the Isaac Newton Group of Telescopes
%
}\\
\vspace*{1cm}
%
Balcells, M.$^{1,2,3}$
%
}\\
\vspace*{0.5cm}
%
$^{1}$
Isaac Newton Group, 38700 Santa Cruz de La Palma, Spain\\
$^{2}$
Instituto de Astrof\'{\i}sica de Canarias, 38200 La Laguna, Tenerife, Spain\\
$^{3}$
Universidad de La Laguna, 38200 La Laguna, Tenerife, Spain
%
\end{flushleft}
%
\markboth{
WEAVE and HARPS3 at ING
}{ 
%
Balcells, M.
%
}
\thispagestyle{empty}
\vspace*{0.4cm}
\begin{minipage}[l]{0.09\textwidth}
\ 
\end{minipage}
\begin{minipage}[r]{0.9\textwidth}
\vspace{1cm}
\section*{Abstract}{\small
%
The Isaac Newton Group of Telescopes is completing a strategic change for the scientific use of its two telescopes, the 4.2-m William Herschel Telescope (WHT) and the 2.5-m Isaac Newton Telescope (INT). After more than 30 years operating as multi-purpose telescopes, the telescopes will soon complete their shift to nearly-single instrument operation dominated by large surveys. 

At the WHT, the WEAVE multi-fibre spectrograph is being commissioned in late 2022. Science surveys are expected to launch in 2023. 30\% of the available time will be offered in open time. For the INT, construction of HARPS-3, a high-resolution ultra-stable spectrograph for extra-solar planet studies, is underway, with deployment planned for late 2024. The INT itself is being modernised and will operate as a robotic telescope. An average of 40\% of the time will be offered as open time. 

The ING will maintain its student programme. Plans call for moving student work from the INT to the WHT once the INT starts operating robotically. 

%
\normalsize}
\end{minipage}
%
%
%
\section{Introduction \label{intro}}
Since the mid 1980's, the Isaac Newton Group\footnote{www.ing.iac.es} (ING) operates the 4.2-m William Herschel Telescope (WHT) and the 2.5-m Isaac Newton Telescope (INT) at the Observatorio Roque de los Muchachos on the Canarian island of La Palma. The telescopes have provided front-line multi-instrument observing capabilities to the ING astronomical communities of the UK, Spain and the Netherlands. Instrumentation comprised facility instruments (most recently ISIS, LIRIS, ACAM, PF, AF2, WYFFOS, INGRID, LDSS, NAOMI, OASIS, TAURUS-II, WFC, IDS), as well as powerful visiting instruments for science or for technology demonstration (AOLI, CANARY, DIPOL, GASP, GHaFaS, HiPERCAM, iQuEyE, LEXI, PAUCAM, PN.S, SPIFS, CIRPASS, CIRSI, ExPo, INTEGRAL, MAOJCam, PLANETPOL, PWFS, SAURON, ULTRACAM, Stereo-SCIDAR, Sodium Profiler). 

In 2010, following a strategic review and community consultation\footnote{https://www.ing.iac.es/about-ING/strategy/},  and following the Astronet Roadmap for European Astronomy \cite{astronetroadmap09} and the European Telescope Strategy Review Committee 2010 report \cite{etsrc10}, ING started making steps to allow the WHT to become a powerful spectroscopic survey facility. This transformation had been announced at previous SEA conferences\cite{balcellsSEA2015}. 

This paper provides a much needed update, after the pandemia years, coinciding with the exciting times when the WHT is starting to collect sky light with its new instrumentation.

\begin{figure}[htbp]
\begin{center}
\includegraphics[scale=1.1]{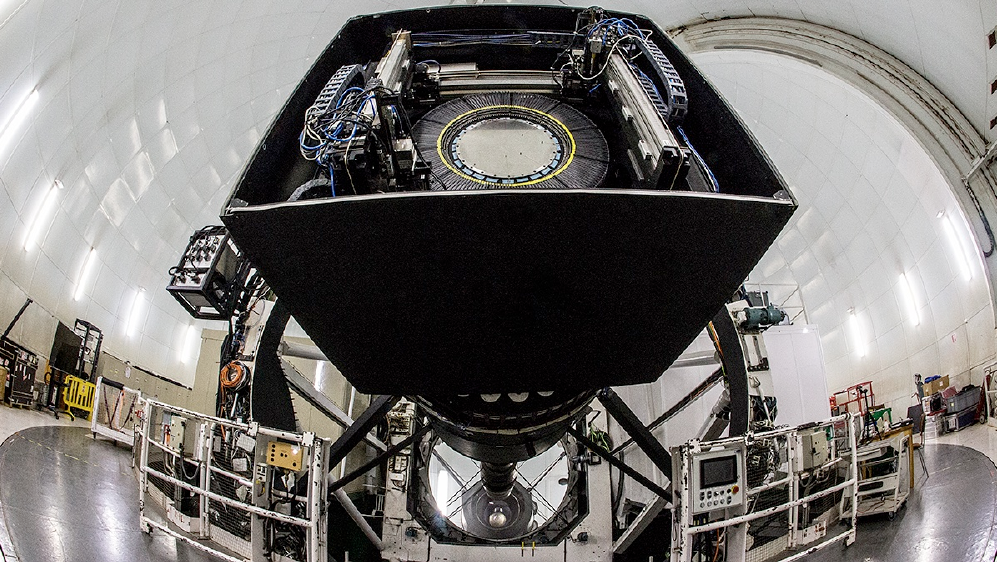}
\caption{The WHT prime focus equiped with the WEAVE fibre positioner. Photo credit: J. M\'endez, ING.}
\label{WEAVEonWHT}
\end{center}
\end{figure}

\section{The new WHT}\label{newWHT}

With a strategic view focused on science from massive spectroscopic surveys doable on a 4-m class telescope, there were clear design drivers for new instrumentation\cite{balcellsSPIE2010}. The WEAVE instrument (Fig.~\ref{WEAVEonWHT}) is the result of that strategy. Three main elements comprise the new WHT:

\begin{description}
\item[Prime-focus corrector.] A new optical corrector delivers a corrected field of view (FOV) of 2 degree (40 cm) diameter, with good transmission from 360 nm to 1000 nm. 
\item[Fibre positioner.] A pick-and-place system, based on the 2dF system at the AAT, employs two robots to place up to 960 fibres on the focal plane. In addition to the standard single-fibre multi-object mode, the instrument comprises 20 fully-deployable mini-integral-field units (mIFU) of 10 arcsec FOV, as well as a monolithic integral-field unit (LIFU) with field of view 90 arcsec. A total of 3243 fibres transmit the light from the  MOS system, the mIFU system and the LIFU to the spectrograph. 
\item[Spectrograph. ] A new two-arm spectrograph provides spectroscopy at resolving power of either 5,000 (LR) or 20,000 (HR). Wavelength coverage at LR is the entire optical range, 366 -- 959 nm.

\end{description}

\subsection{WEAVE highlights}\label{WEAVE}

The WEAVE instrument has been described elsewhere  \cite{daltonSPIE2012}\cite{daltonSPIE2014}\cite{terrettSPIE2014}\cite{balcellsEAS2014}. The definitive description of the as-built instrument, at the time of integration at the telescope, pre-commissioning, is given in \cite{jinSurveyPaper2023}. And an updated description is being maintained at the ING web site\footnote{https://www.ing.iac.es/astronomy/instruments/weave/weaveinst.html}. 

We direct the reader to the above references, and here just note the three input modes: multi-object mode (MOS), the large-integral field mode (LIFU) and the mini-integral field mode (mIFU). We discuss WEAVE's main highlights in the global context of multi-object spectroscopy instruments. 

\begin{table}[htp]
\caption{WEAVE highlights}
\begin{center}
\begin{tabular}{|l|l|}
\hline
Parameter 			& Interest	\\ \hline
MOS lowres $R=5,000$    & $\delta v \sim 3\,\mathrm{km\,s^{-1}}$ @ $V\sim20$,  match to Gaia $v_\mathrm{T}$ \\
MOS highres $R=20,000$ & Improved continuum for line strengths  \\
MOS, mIFU FOV $\sim 2\,\mathrm{deg}$           & High multiplex \\
LIFU highres $R=10,000$ & Resolving vertical vel. disp. of galaxy disks \\ 
LIFU FOV $\sim 90\,\mathrm{arcsec}$     & Evolution of PPAK \\
20 mIFU, FOV $\sim 11\,\mathrm{arcsec}$ & MANGA but on a 4-m, $R=5,000$ \\
End-to-end pipeline                                     & Science-ready data \\ 
Offered for surveys and open time               & Accommodates both large and small projects \\   \hline
\end{tabular}
\end{center}
\label{WEAVEhighlights}
\end{table}%

When compared to other MOS instruments on 4-m class telescopes, WEAVE shines in a number of aspects. The more salient of them are noted in Table~\ref{WEAVEhighlights}:

\begin{itemize}
\item In its default, low-resolution mode, WEAVE's resolving power $R=5000$ is high among high-multiplex MOS instruments built for galactic and extra-galactic science, and makes WEAVE ideal for Milky Way dynamics and archaeology, as it provides stellar radial velocities with accuracies similar to those of tangential velocity data from Gaia. 
\item In its high-resolution mode, WEAVE's resolving power $R=20,000$, when combined with its high multiplex, represents a significant step forward for stellar line strength determinations, as continua adjacent to the lines are less affected by instrumental broadening. 
\item The MOS FOV, 2 degree diameter, is unique in its mIFU mode and in the high-resolution MOS. 
\item The LIFU high-resolution configuration, which delivers R=10,000 spectra, can resolve the vertical velocity structure of galaxy disks. 
\item The LIFU FOV, which was patterned after the PPAK LIFU\cite{kelzPPAK2006}, has 50\% higher FOV, and higher wavelength coverage ($3660-9590~\AA$), and represents a true next-generation large IFU for the study of extended sources. 
\item The mIFU FOV matches the smaller end of the distribution of FOV for the MANGA units\cite{bundyMANGA2015}. With their wide wavelength coverage and with spectral resolutions  2.5[10] times higher than the SDSS-III spectrographs in the low[high] WEAVE resolutions, the WEAVE mIFU's will be powerful tools for low-mass galaxies and for compact star-forming regions in our Galaxy. 
\item Data will be delivered fully reduced and containing a range of science-ready products. 
\end{itemize}

\subsection{WEAVE surveys}\label{surveys} 

Eight surveys (Table~\ref{TableWEAVEsurveys}) have been approved for execution over 5 years of WEAVE operation; they will be allocated 70\% of the available time on WEAVE. As of this writing, the most comprehensive description of the surveys is given in \cite{jinSurveyPaper2023}. For updates and contact points for the surveys and for WEAVE science, the reader is directed to the WEAVE web site\footnote{https://ingconfluence.ing.iac.es/confluence/display/WEAVEDEV/WEAVE\%3A+The+Science}.

\begin{table}[htp]
\caption{The eight WEAVE surveys}
\begin{center}
\begin{tabular}{|ll|ll|}
\hline
& Title                      & & Title \\ \hline
GA &Galactic Archaeology      & SCIP &Stellar, Circumstellar and Interstellar Physics \\
WC &Galaxy Clusters\cite{cornwellClusters2022}              & StePS &Stellar Populations At Intermediate Redshift\cite{iovinoStePS2023} \\
WA &WEAVE Apertif \cite{hessApertif2020}               & WL &WEAVE LOFAR\cite{smithWL2016} \\
WQ &WEAVE QSO\cite{kraljicWQ2022}                   & WD &WEAVE White Dwarfs \\ \hline

\end{tabular}
\end{center}
\label{TableWEAVEsurveys}
\end{table}%

\subsection{Using the WHT: surveys and open time}\label{usingWHT}

ING is offering time on WEAVE for large surveys as well as through the ING national time-allocation committees. The International Time programme of the Canarian Observatories\footnote{https://www.iac.es/en/observatorios-de-canarias/observing-time/observing-time/international-time-programme} provides an additional means of obtaining up to 15 WEAVE nights per year in addition to time on any of the other ORM telescopes. 

WEAVE is due to start commissioning in the fall of 2022. We anticipate LIFU science verification observations in early 2023, and aim for starting science in the middle of 2023. 

\section{The third life of the Isaac Newton Telescope}\label{INT}

ING is in the middle of transforming the 2.5-m Isaac Newton Telescope (INT) by installing HARPS3\footnote{https://www.terrahunting.org/harps3.html} \cite{thompsonSPIE2016}, an enhanced version of the HARPS and HARPS-N spectrographs aimed at achieving 10 cm\,s$^{-1}$ radial velocity precision on nearby stars to search for Earth-like planets. HARPS3 differs from its predecessors by a stabilised beam feed and a polarimetric sub-unit\cite{dorvalSPIE2018} which will provide a powerful tool for characterising stellar activity. The Terra Hunting Experiment (THE) consortium, P.I. Didier Queloz, is building HARPS3 and making it available in exchange for $\sim$50\% of every night over 10 years; the remaining time will be offered as open time through the usual national time allocation channels. The consortium is also modernising the INT, which will become a robotic telescope. This will be the third encarnation of the venerable INT, after its first installation in Herstmonceux in 1967 and it re-deployment in La Palma in 1982. 

With this transformation, we expect the INT will provide the UK, ES and NL astronomical communities with a much needed tool for extra-solar planet science. 

Current plans call for the robotic INT to be commissioned during the summer of 2024, and for HARPS3 to start scientific operations before the end of 2024. 

\section{Opportunities for students}\label{students}

ING will continue to welcome 4 to 6 students for year-long stays at the ING. Building on the success of this highly-demanded programme, we are hoping to increase the number of student positions in the near future. 
 
When the INT closes down for reforms and becomes a robotic telescope, students will dominantly work at the WHT, partaking in WEAVE survey and open-time observations. This will open up opportunities for the students to become familiar with the execution of large projects and to develop expertise in WEAVE data. 

%
%
\small  
%
\section*{Acknowledgments}   
%
The Isaac Newton Group of Telescopes is operated on behalf of the UK Science and Technology Facilities Council (STFC) the Nederlanse Organisatie voor Wetenschappelijk Onderzoek (NWO), and the Instituto de Astrof\'{\i}sica de Canarias (IAC). 

WEAVE construction was funded from generous grants from STFC, NWO, Spanish Science Ministry, the French CNRS, and the Italian INAF. Additional contributions were received from Konkoly Observatory in Hungary, INAOE in Mexico, and PI grants from Lund Observatory, IAP Potsdam, MPIA Heidelberg. 

The HARPS3 instrument is being built by the Terra Hunting Experiment Consortium led by University of Cambridge Cavendish Laboratory.

%

%
\end{document}